\documentstyle[preprint,aps,epsf]{revtex}
\def\lessim{\mathrel {\vcenter {\baselineskip 0pt \kern 0pt
\hbox{$<$} \kern 0pt \hbox{$\sim$} }}}
\def\gessim{\mathrel {\vcenter {\baselineskip 0pt \kern 0pt
\hbox{$>$} \kern 0pt \hbox{$\sim$} }}}
\def \rightdownarrow
 {\kern.4em {\raise1.75ex\hbox{$\Bigl |$}$\kern-0.27em{\longrightarrow}$}}
\def \mrightdownarrow
 {\kern.4em {\raise1.25ex \hbox{$|$} \kern-0.27em{
                                                \longrightarrow}}}

\def\GeVc{GeV$\!/c$}
\def\GeVcc{GeV$\!/c^2$}
\def\mGeVcc{{\rm GeV\!/c^2}}

\def\mBR{{\cal B}}

\def\mMEt{\not\kern-.35em {E_T}}
\def\MEt{\hbox{$\mMEt$}}

\def\etal{{\it et al.}}
\def\Wprime{W^\prime}
\def\MWprime{M_{W^\prime}}
\def\MnuR{M_{\nu_R}}
\def\tbbar{t\bar{b}}
\def\ttbar{t\bar{t}}
\def\bbbar{b\bar{b}}
\font\eightit=cmti8
\def\r#1{\ignorespaces $^{#1}$}
\begin{document}
\title{
\begin{flushright}
{
\small
CDF/PUB/EXOTIC/PUBLIC/5927 \\
\today
}
\end{flushright}
Search for a $W^\prime$\ Boson Decaying to a Top and Bottom Quark Pair in
1.8 TeV $p\bar{p}$\ Collisions }
\author{
\font\eightit=cmti8
\def\r#1{\ignorespaces $^{#1}$}
\hfilneg
\begin{sloppypar}
\noindent
D.~Acosta,\r {14} T.~Affolder,\r {25} H.~Akimoto,\r {51}
M.~G.~Albrow,\r {13} D.~Ambrose,\r {37}   
D.~Amidei,\r {28} K.~Anikeev,\r {27} J.~Antos,\r 1 
G.~Apollinari,\r {13} T.~Arisawa,\r {51}
% ...
\end{sloppypar}
\vskip .026in
\begin{center}
(CDF Collaboration)
\end{center}

\vskip .026in
\begin{center}
\r 1  {\eightit Institute of Physics, Academia Sinica, Taipei, Taiwan 11529, 
Republic of China} \\
\r 2  {\eightit Argonne National Laboratory, Argonne, Illinois 60439} \\
\r 3  {\eightit Istituto Nazionale di Fisica Nucleare, University of Bologna,
I-40127 Bologna, Italy} \\
\r 4  {\eightit Brandeis University, Waltham, Massachusetts 02254} \\
\r 5  {\eightit University of California at Davis, Davis, California  95616} \\
\r 6  {\eightit University of California at Los Angeles, Los 
Angeles, California  90024} \\ 
\r 7  {\eightit University of California at Santa Barbara, Santa Barbara, California 
93106} \\ 
\r 8 {\eightit Instituto de Fisica de Cantabria, CSIC-University of Cantabria, 
39005 Santander, Spain} \\
\r 9  {\eightit Carnegie Mellon University, Pittsburgh, PA  15218} \\
\r {10} {\eightit Enrico Fermi Institute, University of Chicago, Chicago, 
Illinois 60637} \\
\r {11}  {\eightit Joint Institute for Nuclear Research, RU-141980 Dubna, Russia}
\\
\r {12} {\eightit Duke University, Durham, North Carolina  27708} \\
\r {13} {\eightit Fermi National Accelerator Laboratory, Batavia, Illinois 
60510} \\
\r {14} {\eightit University of Florida, Gainesville, Florida  32611} \\
\r {15} {\eightit Laboratori Nazionali di Frascati, Istituto Nazionale di Fisica
               Nucleare, I-00044 Frascati, Italy} \\
\r {16} {\eightit University of Geneva, CH-1211 Geneva 4, Switzerland} \\
\r {17} {\eightit Glasgow University, Glasgow G12 8QQ, United Kingdom}\\
\r {18} {\eightit Harvard University, Cambridge, Massachusetts 02138} \\
\r {19} {\eightit Hiroshima University, Higashi-Hiroshima 724, Japan} \\
\r {20} {\eightit University of Illinois, Urbana, Illinois 61801} \\
\r {21} {\eightit The Johns Hopkins University, Baltimore, Maryland 21218} \\
\r {22} {\eightit Institut f\"{u}r Experimentelle Kernphysik, 
Universit\"{a}t Karlsruhe, 76128 Karlsruhe, Germany} \\
\r {23} {\eightit High Energy Accelerator Research Organization (KEK), Tsukuba, 
Ibaraki 305, Japan} \\
\r {24} {\eightit Center for High Energy Physics: Kyungpook National
University, Taegu 702-701; Seoul National University, Seoul 151-742; and
SungKyunKwan University, Suwon 440-746; Korea} \\
\r {25} {\eightit Ernest Orlando Lawrence Berkeley National Laboratory, 
Berkeley, California 94720} \\
\r {26} {\eightit University College London, London WC1E 6BT, United Kingdom} \\
\r {27} {\eightit Massachusetts Institute of Technology, Cambridge,
Massachusetts  02139} \\   
\r {28} {\eightit University of Michigan, Ann Arbor, Michigan 48109} \\
\r {29} {\eightit Michigan State University, East Lansing, Michigan  48824} \\
\r {30} {\eightit Institution for Theoretical and Experimental Physics, ITEP,
Moscow 117259, Russia} \\
\r {31} {\eightit University of New Mexico, Albuquerque, New Mexico 87131} \\
\r {32} {\eightit Northwestern University, Evanston, Illinois  60208} \\
\r {33} {\eightit The Ohio State University, Columbus, Ohio  43210} \\
\r {34} {\eightit Osaka City University, Osaka 588, Japan} \\
\r {35} {\eightit University of Oxford, Oxford OX1 3RH, United Kingdom} \\
\r {36} {\eightit Universita di Padova, Istituto Nazionale di Fisica 
          Nucleare, Sezione di Padova, I-35131 Padova, Italy} \\
\r {37} {\eightit University of Pennsylvania, Philadelphia, 
        Pennsylvania 19104} \\   
\r {38} {\eightit Istituto Nazionale di Fisica Nucleare, University and Scuola
               Normale Superiore of Pisa, I-56100 Pisa, Italy} \\
\r {39} {\eightit University of Pittsburgh, Pittsburgh, Pennsylvania 15260} \\
\r {40} {\eightit Purdue University, West Lafayette, Indiana 47907} \\
\r {41} {\eightit University of Rochester, Rochester, New York 14627} \\
\r {42} {\eightit Rockefeller University, New York, New York 10021} \\
\r {43} {\eightit Instituto Nazionale de Fisica Nucleare, Sezione di Roma,
University di Roma I, ``La Sapienza," I-00185 Roma, Italy}\\
\r {44} {\eightit Rutgers University, Piscataway, New Jersey 08855} \\
\r {45} {\eightit Texas A\&M University, College Station, Texas 77843} \\
\r {46} {\eightit Texas Tech University, Lubbock, Texas 79409} \\
\r {47} {\eightit Institute of Particle Physics, University of Toronto, Toronto
M5S 1A7, Canada} \\
\r {48} {\eightit Istituto Nazionale di Fisica Nucleare, University of Trieste/\
Udine, Italy} \\
\r {49} {\eightit University of Tsukuba, Tsukuba, Ibaraki 305, Japan} \\
\r {50} {\eightit Tufts University, Medford, Massachusetts 02155} \\
\r {51} {\eightit Waseda University, Tokyo 169, Japan} \\
\r {52} {\eightit University of Wisconsin, Madison, Wisconsin 53706} \\
\r {53} {\eightit Yale University, New Haven, Connecticut 06520} \\
\end{center}

%
%
}
\draft
\address{}
\maketitle
%
%

\begin{abstract}
We report the results of a search for a $W^\prime$\ boson produced 
in $p\bar p$\ collisions  at a center-of-mass energy of 1.8 TeV
using a $106~\hbox{pb}^{-1}$\ data sample recorded by the Collider Detector at
Fermilab. We observe no significant excess of events above background 
for a 
$\Wprime$\ boson decaying to a top and bottom quark pair, with the top quark
subsequently decaying into a semileptonic final state.
These data allow us to set limits on the rate of
$W^\prime$\ boson production and decay.
In a model where this boson would mediate interactions
involving a massive right-handed neutrino ($\nu_R$) and has
Standard Model strength couplings, we exclude a
$W^\prime$\ boson with mass between 225 and 
$536~\hbox{GeV}\!/c^2$\ at 95\% confidence level for $\MWprime \gg \MnuR$\
and between 225 and 
$566~\hbox{GeV}\!/c^2$\ at 95\% confidence level for $\MWprime < \MnuR$.\
\end{abstract}

%
%

%
The search for additional forces in nature have focused on identifying
particle physics phenomena not predicted by the 
strong, electromagnetic and weak forces.
These are described by the Standard Model using a local gauge
theory that accounts for each interaction using a vector boson force
carrier
\cite{ref: EWK Reference}.   
Evidence for a new force could come from observation of
the  corresponding force carrier.
There are a number of extensions to the Standard Model that predict the
existence of a new charged vector boson, generically known as a
$\Wprime$\ boson.  The most common extensions are left-right symmetric 
\cite{ref: LR Symmetric Models},
in that they presume that the $\Wprime$\ boson mediates right-handed
interactions, in the same way that the Standard Model $W$\ boson mediates
only left-handed interactions.  

Previous searches for new charged vector bosons with couplings to quarks
and leptons have been reported. 
These searches have set model dependent
limits on the new boson mass as well as limits on cross section times 
branching ratio. 
A comprehensive set of searches has been performed at the Fermilab
Tevatron Collider.  
Searches using the decay mode 
$\Wprime \rightarrow e\nu_e$\ exclude a $\Wprime$\ boson with mass
$< 754~\mGeVcc$\ at 95\%\ CL 
\cite{ref: Wprime searches electron CDF,ref: Wprime searches electron D0}, 
while similar  searches considering the decay mode
$W^\prime\rightarrow\mu\nu_\mu$\ have excluded a $W^\prime$\ boson with
mass $<660~\mGeVcc$\ at 95\%\ CL \cite{ref: Wprime searches muon}. 
The most stringent single limit comes from a search combining both of
these leptonic channels and excludes a $\Wprime$\ boson with mass
$< 786~\mGeVcc$\ at 95\%\ CL \cite{ref: Wprime searches electron CDF}. 
These mass limits all assume that the new vector boson's couplings to
leptonic final states will be given by the Standard Model, with the
additional assumption that
the mass of the neutrino produced in the leptonic decay of the $\Wprime$\ is
much less than the mass of the $\Wprime$\ boson itself.   
With these assumptions, the predicted total width of the boson 
increases linearly with
$\MWprime$, where $\MWprime$\ is the mass of the boson. 
A search that
avoids any assumptions regarding the neutrino mass has
involved the decay mode $\Wprime \rightarrow q \bar{q}^\prime$\ where the
quarks are observed as high-energy jets, but is background-limited and
only excludes $\Wprime$\ bosons with 
$300 < \MWprime < 420~\mGeVcc$\ at 95\%\ CL \cite{ref: Wprime searches
hadronic}.
Indirect searches studying, for example, the Michel spectrum in
$\mu$\ decay have resulted in more model-independent limits with less 
sensitivity \cite{ref: indirect searches}.

In this letter, we present the results
of a new search for a 
$W^\prime$\ boson decaying to a top quark-bottom quark pair, {\it i.e.}
$\Wprime\rightarrow \tbbar$.  Although this search is only
sensitive to
$\Wprime$\ bosons with mass above the $t\bar{b}$\ kinematic threshold of
approximately 200~\GeVcc, it is relatively free of background 
compared to the $W^\prime\to q\bar q^\prime$\ decay mode 
because of the signature from the top quark decay
$t\rightarrow W b$.
Furthermore, the interpretation of the data is less sensitive to
assumptions regarding the right-handed neutrino sector, since
we avoid having to make assumptions regarding the nature of
the right-handed neutrino or the leptonic couplings of the $W^\prime$
boson \cite{ref: Rosner PRD}.
We use a data sample of
$106\pm 4~\hbox{pb}^{-1}$\ of 1.8 TeV
$p\bar p$\ collisions recorded by the Collider Detector at Fermilab (CDF)
detector during 1992-95.  

The CDF detector is described in detail elsewhere
\cite{ref: CDF Detector}. 
The detector has a charged particle tracking
system immersed in a 1.4~T solenoidal magnetic field, which is coaxial
with the $p\bar p$\ beams. 
The tracking system consists of silicon strip
detectors and drift chambers that measure particle momentum with an
accuracy of 
$
\sigma_{p_T}/p_T = \sqrt{ (0.0009 p_T)^2 + (0.0066)^2},
$
where $p_T$\ is the momentum of the charged particle measured in \GeVc\
transverse to the $p\bar p$\ beamline.
The tracking system is surrounded by segmented electromagnetic and
hadronic calorimeters measuring the flow of energy associated with
particles that interact hadronically or electromagnetically out to 
$|\eta|$\ of 4.2~\cite{ref: Coord-system}.
Electron candidates with $|\eta|<1.0$\ are identified using the pattern of
energy distribution in the electromagnetic and hadronic calorimeters and
the presence of a charged track consistent with the calorimeter
information. A set of charged particle detectors outside the calorimeter
is used to identify muon candidates with $|\eta|<1.0$.  

Because of the good background rejection of the electron and muon
identification systems in the CDF detector, we choose to search for those
events that are consistent with 
$\Wprime\rightarrow \tbbar$\ with the top quark decaying to final
states including 
either 
$e \nu_e b$\ or $\mu \nu_\mu b.$\  The
primary selection criteria is identical to an earlier study
searching for single top quark production \cite{ref: CDF single
top}.  Candidate events 
are
identified in the CDF trigger system by the
requirement of at least one electron or muon candidate with
$p_T > 18$~\GeVc.  
The event sample 
is
subsequently
refined after full event reconstruction by requiring a well-identified
electron or muon candidate with $p_T>20$~\GeVc\ and by requiring
that the missing transverse energy in the event, $\MEt$, be greater than
20~GeV.  We 
reject 
events that 
are
identified as dilepton 
candidates arising from top quark pair production ($\ttbar$) \cite{ref:
CDF ttbar dilepton}, and 
reject
other dilepton candidates as described
in Ref.~\cite{ref: CDF single top}. 
For example, candidate events with two well-identified electron or muon
candidates consistent with coming from top quark decay were rejected.
To select events with at least two
bottom quark candidates, we 
require
either two or three jets with
transverse energy $E_T > 15$~GeV and $|\eta|<2.0$, where the jets 
are
defined using 
a fixed-cone clustering algorithm employing a cone-size
of $R\equiv\sqrt{(\Delta\eta)^2 + (\Delta\phi)^2}=0.4$.  The jet
transverse energies 
are
subsequently corrected with an algorithm that
accounts for the effects of jet fragmentation, calorimeter
non-uniformities and energy flow from the rest of the 
event~\cite{ref: jet correction}.  
We
require 
that at least one of the jets 
be identified
as a $b$\ quark candidate using displaced secondary vertex
information from the silicon vertex detector \cite{ref: CDF bbar new
particle search}.  This selection 
results 
in 57 candidate events.

We 
use
a PYTHIA Monte Carlo calculation \cite{ref: PYTHIA}\ and a
CDF detector simulation to determine the expected number of
candidate events we would observe in this data sample as a function of
$\Wprime$\ boson mass.  We 
require
the $\Wprime$\ boson to have a right-handed
coupling to the $\tbbar$\ final state, we set the top quark mass to
175~\GeVcc\ and we 
assume that the top quark always decays to a $Wb$\
final state.  
We expect negligible signal yield differences between right-handed and
left-handed couplings.
We 
use
the MRS(G) parton distribution functions \cite{ref:
MRSG PDF}\ to model the momentum distribution of the initial state
partons.  
We use a next-to-leading-order calculation to estimate the production cross 
section, as the next-to-leading order contributions are substantial
\cite{ref: K factor}; the increase in cross section over the 
leading-order prediction ranges from a factor of 1.50 at $\MWprime
=225~\mGeVcc$\ to 1.26 at $\MWprime=600~\mGeVcc$.  The 
efficiency times acceptance in both the electron and muon channels for our
event selection is approximately 9\%\ for $\MWprime=225~\mGeVcc$,
increases to 12\%\ for
$\MWprime=300~\mGeVcc$\ and is approximately constant for masses up to
$600~\mGeVcc$.  
The corresponding efficiency times acceptance for the
$\tau$\ lepton channel, where this lepton decays to an energetic muon or 
electron, is approximately a factor of six to ten smaller.  
We will not attempt to interpret our
data for masses $\MWprime < 225$~\GeVcc\ as the cross section calculation
and the acceptance calculation 
become increasingly uncertain as one nears the kinematic
threshold for the decay $W^\prime \rightarrow \tbbar$.  
The production cross section times branching ratio and the expected number of
signal events  as a function of
$\MWprime$\ are shown in Table~\ref{tab: Expectations}.  Over a wide-range
of $\Wprime$\ boson masses, we would expect to see significant numbers of
events contributing to our candidate sample.  

We identified three
sources that comprise the dominant background contributions to this search:  
the pair production 
of top quarks ($\ttbar$), single top quark production and the
associated QCD production of $W$\ bosons with one or more heavy quarks
($W\bbbar$\ and $Wc$\ where $c$\ is the charm quark).
We have investigated other
possible background sources and find them to be individually 
insignificant.
We estimate the background contribution from $\ttbar$\ production
by employing a PYTHIA Monte Carlo calculation of this process and a
detector simulation.  Using the predicted $\ttbar$\ production
cross section of $5.1\pm0.9$~pb \cite{ref: ttbar production sigma}, we
estimate the $\ttbar$\ background to be $15.0\pm4.0$\ observed events.  We
use the same methods as in our measurement of single top quark
production search \cite{ref: CDF single top}\ to estimate the single top
quark contribution to our candidate sample to be $3.9\pm0.9$\ observed events. 
The largest single background contribution comes from the associated QCD
production of $W$\ bosons with heavy quarks.  
We 
employ 
the technique
described in an earlier report \cite{ref: CDF bbar new particle search}\ to estimate
these, taking into account the different event selection requirements,
and 
find
a total expected background contribution of
$15.6\pm3.0$\ events.  Other sources of background, including events not
containing a heavy quark jet, dilepton final states and events with
misidentified lepton candidates are predicted to give rise to 
$13.6\pm2.9$\ events.  We thus expect $48\pm6$\ candidate events from
background processes.  This is in reasonable agreement with the 57 candidate
events observed, and we conclude we have no significant evidence for $\Wprime$\
boson production.  

\begin{figure}
\begin{center}
\leavevmode
\hbox{%
 \epsfxsize=5.4in
 \epsffile{
 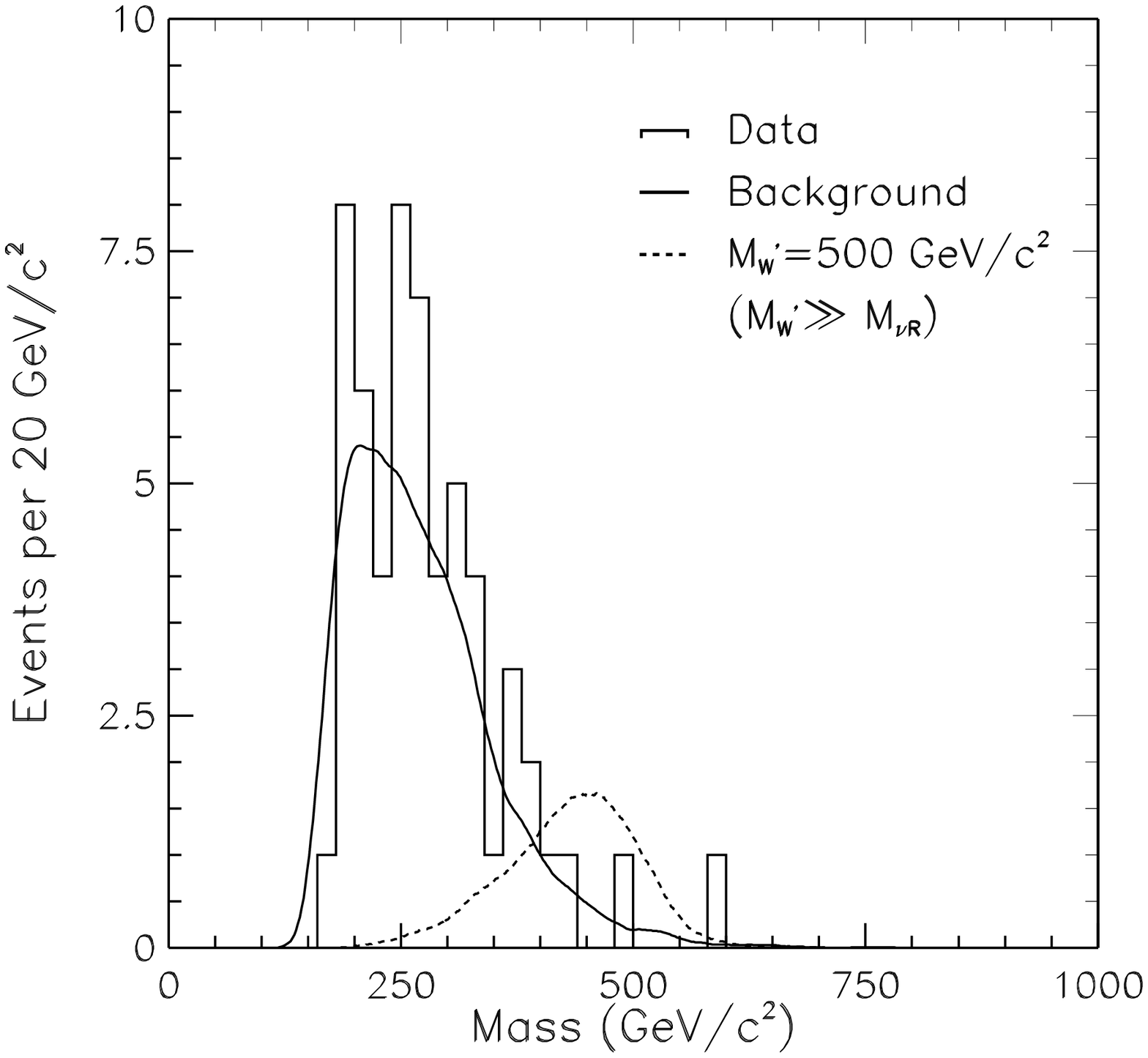
 }
}
\end{center}
\caption
{The $W\bbbar$\ mass spectrum of the candidate events after constraining
the lepton-neutrino invariant mass to the $W$\ boson mass.   The 
distribution expected from the production of a 
$W^\prime$\ boson with a mass of 500~\GeVcc\ is illustrated by the dashed
curve.  The distribution expected from the background processes is
shown by the solid curve, which is normalized to the expected total
background rate.}
\label{fig: expected mass}
\end{figure}

To set a limit on the $W^\prime$ mass, we employ the invariant mass 
distribution 
of the  $W\bbbar$\ final state
as that provides more information about possible $W^\prime$\ production
than the number of candidate events alone.  We reconstruct
the momentum of the neutrino along the beam axis ($p_z$) by constraining
the invariant mass of the lepton-neutrino pair to equal the
$W$\ boson mass of 80.22~\GeVcc \cite{ref: PDG}, which results
in a quadratic constraint on $p_z$.  This generally provides two
solutions, and we select the solution with the smaller value of
$|p_z|$ as that is more likely correct given the 
central nature of the production mechanism of this very heavy state.  
If the solution has an imaginary component, we use only the
real component.  The resulting $W\bbbar$\ mass distribution for our 57 candidate
event sample is shown in Fig.~\ref{fig: expected mass}\ and is compared with the
expected mass distribution for a 
$\Wprime$\ boson with $\MWprime=500~\mGeVcc$\ .  We also show the mass
distribution expected from the sum of the background processes.  

\begin{figure}
\begin{center}
\leavevmode
\hbox{%
 \epsfxsize=5.4in
 \epsffile{
 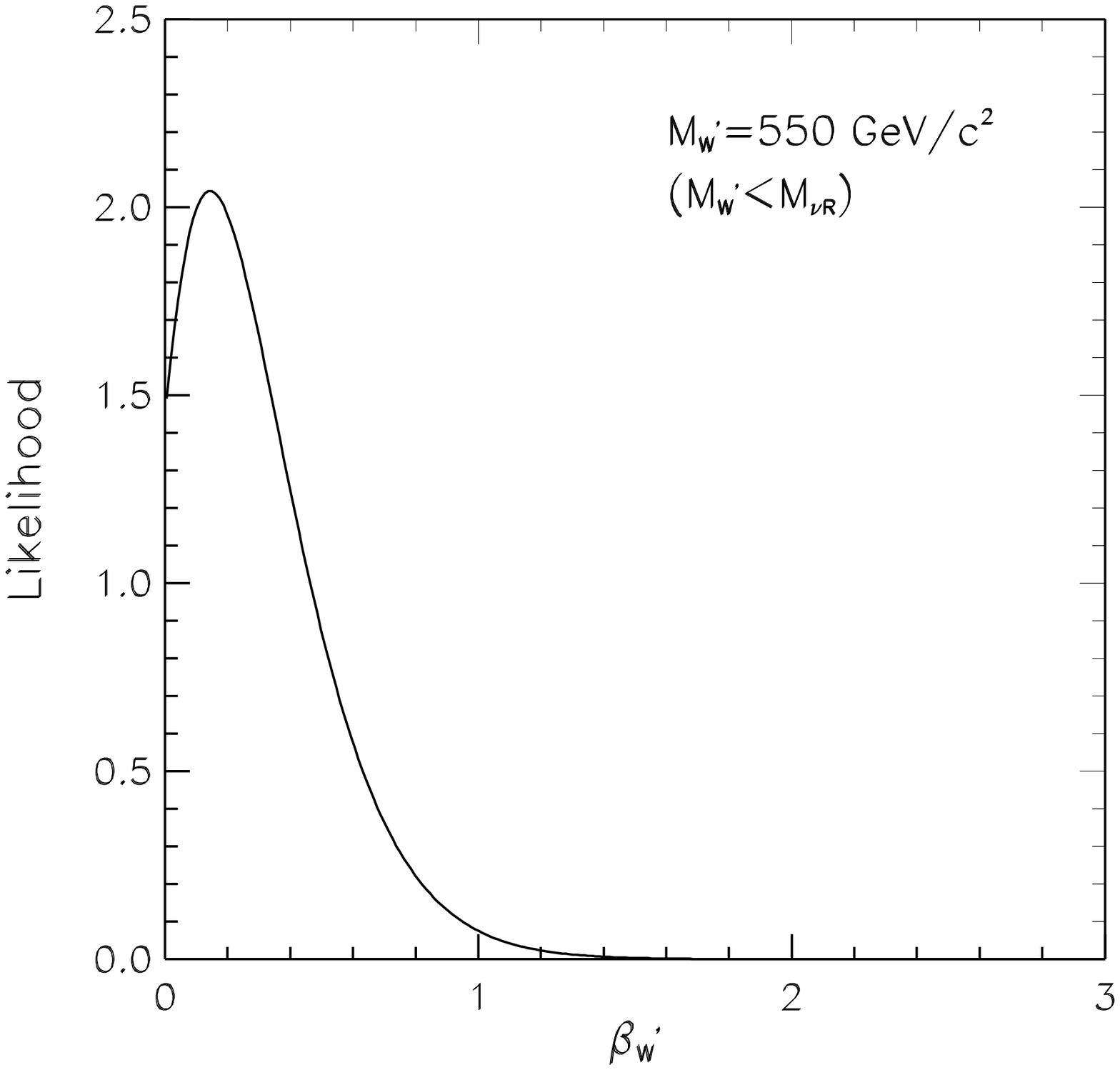
 }
}
\end{center}
\caption{
The dependence of the likelihood on $\beta_{W^\prime}$\ with the assumption
of $\MWprime=550~\mGeVcc$\ and 
$\MWprime < \MnuR$.  The likelihood plots with other assumptions are similar.} 
\label{fig: likelihood plot}
\end{figure}

To estimate the
size of the potential signal contribution, we perform an unbinned maximum
likelihood fit to both the number of observed events and the
observed mass distribution, allowing for both a signal and background
contribution for different values of
$\MWprime$\ ranging from 225 to 600~\GeVcc.  
We use a fitting technique
that is identical to that employed in the recent search for single top
quark production \cite{ref: CDF single top}, where we model the expected
mass distribution as a sum of a signal component with size $\beta_{W^\prime}$, 
and three background components with sizes $\beta_{\ttbar}$, $\beta_{st}$\
and $\beta_{nt}$ for the backgrounds from top quark pair production, 
single top quark production, and sources not containing a top quark,
respectively. These parameters are normalized so that they equal
unity when the fit results in the number of observed events equal to the
number of predicted events from each individual source.   With this
choice of normalization, we can interpret 
\begin{eqnarray}
\beta_{W^\prime} = {{\sigma\cdot\mBR(\Wprime\rightarrow \tbbar)}\over{
\sigma\cdot\mBR(\Wprime\rightarrow \tbbar)_{SM}}},
\end{eqnarray}
where the denominator is the expected production cross section times
branching fraction for the $\Wprime$\ boson assuming Standard Model
couplings.  Since the latter depends on the nature of the
right-handed neutrino, we express our results using the two scenarios
described earlier.   We include in the likelihood Gaussian constraints on
the expected number of events from the three background sources.
The results of the fit are presented in
Table~\ref{tab: fit results}.  An example of the dependence of the likelihood on
$\beta_{W^\prime}$\ is shown in Fig.~\ref{fig: likelihood plot}\ for the case
$\MWprime=550~\mGeVcc$\ and $\MWprime < \MnuR$. 

We set Bayesian 95\%\ CL upper limits on the relative contribution of a
$W^\prime$\ boson by constructing a posterior distribution
$f(\beta_{W^\prime})$\ for each fixed value of $\MWprime$.
First we 
maximize
the
likelihood function for fixed values of
$\beta_{W^\prime}$\ and multiply the resulting function by a flat prior
distribution for $\beta_{W^\prime}$.  We then convolute $f(\beta_{W^\prime})$\ 
with two
Gaussian prior distributions to take into account the systematic uncertainties
that affect the number of expected background or signal events and
the shape of the resulting invariant mass distribution.  The largest
systematic uncertainties arise from our uncertainty in the efficiency
for $b$\ quark tagging (11\%), our understanding of the lepton
selection efficiency (10\%) and on the parton distribution functions
(between 4 and 11\%).  We are also sensitive to the 
value of the top quark mass
near kinematic threshold; its current uncertainty of
$\pm5$~\GeVcc~\cite{ref: top mass uncertainty} results in a systematic 
uncertainty
on the acceptance  and cross section of 15\%\ at $\MWprime=225~\mGeVcc$, 8\%\ at
$\MWprime=250$~\GeVcc, and $\le 4$\%\ for higher masses.  The 
systematic uncertainties from all effects total approximately 20\%
for $\Wprime$\ boson masses ranging from $\MWprime=225~\mGeVcc$\ to $600~\mGeVcc$.
To set a 95\% CL upper
limit on $\beta_{W^\prime}$, we integrate the 
posterior distribution $f(\beta_{W^\prime})$.
A frequentist calculation of this limit yields 
consistent
results.

The results of the fit and this limit-setting procedure are summarized in
Table~\ref{tab: fit results}\ and plotted in Fig.~\ref{fig: upper
limits}.  We find that we can exclude a $\Wprime$\ boson at 95\%\ CL with masses
$225 < \MWprime < 536$~\GeVcc\ for $\MWprime \gg \MnuR$\ and 
$225 < \MWprime < 566$~\GeVcc\ assuming $\MWprime < \MnuR$.   

In summary, we have performed a search for the
production of a new heavy vector gauge boson in 1.8 TeV $p\bar{p}$\
collisions and decaying into the $\tbbar$\ final state.  We see no
evidence for a signal above the expected background contributions.  We use
a fit of the
final state invariant mass distribution to exclude a
$W^\prime$\ boson with 
$225 < \MWprime < 536$~\GeVcc\ for $\MWprime \gg \MnuR$\ and 
$225 < \MWprime < 566$~\GeVcc\ for $\MWprime < \MnuR$.  
This is the first study made of this production process, and we expect
that it will continue to be an effective search signature for
higher mass $\Wprime$\ bosons that might be produced at future higher
energy and higher luminosity colliders.

We thank the Fermilab staff and the technical staff at the
participating institutions for their essential contributions to this
research.  This work is supported by the U.~S.~Department of Energy
and the National Science Foundation; 
the Italian Istituto Nazionale di Fisica Nucleare; 
the Natural Sciences and Engineering Research Council of Canada; 
the Ministry of Education, Culture, Sports, Science and Technology of Japan; 
the National Science Council of the Republic of China; 
the Swiss National Science Foundation; 
the A.~P.~Sloan Foundation;
the Bundesministerium fuer Bildung und Forschung, Germany; 
the Korea Science and Engineering Foundation (KoSEF); 
the Korea Research Foundation; 
and the Comision Interministerial de Ciencia y Tecnologia, Spain.

%

%
%

%
%
\begin{figure}
\begin{center}
\leavevmode
\hbox{%
 \epsfxsize=5.4in
 \epsffile{
 limits_prl.epsi
 }
}
\end{center}
\caption{
The upper limits on the $W^\prime$\ boson production cross section as a
function of the $W^\prime$\ boson mass. Limits are shown for the case
$\MWprime \gg \MnuR$\ (solid) and $\MWprime < \MnuR$\ (dashed). 
The intercepts at ${\sigma{\cal B}(W^\prime\to tb)\over \sigma{\cal B}
(W^\prime\to tb)_{SM}}=1$ correspond to the 95\% CL limits on 
the $W^\prime$ boson mass with Standard Model strength couplings.}
\label{fig: upper limits}
\end{figure}

\newpage

\begin{table}
\begin{center}
\begin{tabular}{ccccc}
$\MWprime$  &  \multicolumn{2}{c}{$\MWprime \gg \MnuR$} &
\multicolumn{2}{c}{$\MWprime < \MnuR$} \\
(\GeVcc)  &  $\sigma\cdot\mBR(\Wprime\rightarrow \tbbar)$\ (pb) & Events &
$\sigma\cdot\mBR(\Wprime\rightarrow \tbbar)$\ (pb) & Events \\
\hline
225   &   53.4   &  116  &  77.2   &  168 \\
300   &   37.4   &  115 &  52.1  &   161 \\
400   &   13.3   &  43  &  18.0   &  58 \\
500   &   4.38   &  14  &  5.87   &  19  \\
600   &   1.43   &  4.5 &  1.89   &  5.9 \\
\hline
\end{tabular}
\end{center}
\caption{
The production cross section times branching fraction and the
number of expected events for different $\Wprime$\ masses and
different assumptions regarding the right-handed neutrino
sector.
}
\label{tab: Expectations}
\end{table}
\begin{table}
\begin{center}
\begin{tabular}{ccccc}
$\MWprime$  &  \multicolumn{2}{c}{$\MWprime \gg \MnuR$} &
\multicolumn{2}{c}{$\MWprime < \MnuR$} \\
(\GeVcc)  &  
Fit  & Upper Limit &
Fit  & Upper Limit \\
\hline
225 & $0.04^{+0.07}_{-0.04}$ & 0.20 & $0.03^{+0.05}_{-0.03}$ & 0.14 \\
300 & $0.07^{+0.07}_{-0.06}$ & 0.21 & $0.05^{+0.05}_{-0.04}$ & 0.15 \\
400 & $0.09^{+0.13}_{-0.09}$ & 0.38 & $0.06^{+0.09}_{-0.06}$ & 0.27 \\
500 & $0.06^{+0.25}_{-0.06}$ & 0.70 & $0.05^{+0.18}_{-0.05}$ & 0.53 \\
600 & $0.31^{+0.51}_{-0.29}$ & 1.74 & $0.23^{+0.38}_{-0.22}$ & 1.32 \\
\hline
\end{tabular}
\end{center}
\caption{
The fit results for the number of events arising from
$\Wprime$\ production, normalized to the expected number of events for a
given $\Wprime$\ mass, and the Bayesian 95\%\ CL upper limit on this
fraction for the two different assumptions on the mass of the
right-handed neutrino.}
\label{tab: fit results}
\end{table}


\begin{references}
\bibitem{ref: Coord-system}
We use a coordinate system where $\theta$\ is the polar
angle to the proton beam, $\phi$\ is the
azimuthal angle about this beam axis, and
$\eta$ is the pseudorapidity defined as $-\ln\tan(\theta/2)$.  
Missing transverse energy, \MEt, is
defined as the magnitude of $-\sum_i E_T^i \hat{n}_i$, where $\hat{n}_i$\
is a unit vector in the azimuthal plane that points from the beamline to
the $i$th calorimeter tower.
\end{references}
\end{document}